\documentclass[aps,preprint,showpacs,showkeys]{revtex4}
\usepackage{amsmath,graphicx}

\input{tcilatex}

\begin{document}

\title{Electronic Structure and Magnetic Exchange Coupling \\
in Ferromagnetic Full Heusler Alloys }
\author{Yasemin Kurtulus, Richard Dronskowski}
\affiliation{Institut f\"ur Anorganische Chemie, \\
Rheinisch-Westf\"alische Technische Hochschule, \\
52056 Aachen, Germany}
\author{German Samolyuk, Vladimir P.\ Antropov}
\affiliation{Ames Laboratory, \\
Ames, IA, 50011, USA}
\date{\today}

\begin{abstract}
\baselineskip 20pt

\noindent Density-functional studies of the electronic structures and
exchange interaction parameters have been performed for a series of
ferromagnetic full Heusler alloys of general formula Co$_2$MnZ (Z = Ga, Si,
Ge, Sn), Rh$_2$MnZ (Z = Ge, Sn, Pb), Ni$_2$MnSn, Cu$_2$MnSn and Pd$_2$MnSn,
and the connection between the electronic spectra and the magnetic
interactions have been studied. Different mechanisms contributing to the
exchange coupling are revealed. The band dependence of the exchange
parameters, their dependence on volume and valence electron concentration
have been thoroughly analyzed within the Green function technique.
\end{abstract}

\pacs{}
\keywords{Heusler alloys, ferromagnetism, electronic structure, Curie
temperature, exchange parameter}
\maketitle

\section{ Introduction}

\noindent The evolving field of spin-electronics has triggered an increasing
interest in materials with full spin polarization at the Fermi level. Many
of these systems have been predicted by means of electronic band-structure
calculations \cite{deGroot83,Ishida98}, and some of them are in use already
as elements in multi-layered magneto-electronic devices such as magnetic
tunnel junctions \cite{Julliere75,Moodera95} and also as giant
magneto-resistance spin valves \cite{Dieny91}. Promising device candidates
are characterized by a strong spin polarization, by high Curie temperatures
and by a large band gap, too. Among the most popular groups of materials is
the extraordinarily large family of magnetic Heusler alloys \cite{heusler}
which is traditionally considered to be an ideal local-moment system \cite%
{Ishikawa,Hamzic,Kuebler83}. This implies that their exchange couplings can
be described by a Heisenberg Hamiltonian which allows the investigation of
the temperature properties of the magnetic systems within a very simple
concept. It therefore seems that the problem of calculating the exchange
interaction parameters with the help of reliable electronic structure
methods must have a very high priority in this field.

Nonetheless, despite a thorough theoretical understanding of the electronic
structures of many full Heusler alloys (see, for example, Refs.\ \cite%
{Kuebler83,Dederichs,Freeman,Nieminen,Ishida95_1, Ishida95_2,Ishida90}),
only very few publications are dedicated to the discussions of magnetic
exchange interactions in these systems. Noda and Ishikawa \cite{Noda} have
extracted the exchange parameters from the spin-wave spectrum using a
model-like Heisenberg fit. On the theoretical side, K\"ubler, Williams and
Sommers focused on the calculated total-energy differences between the
ferromagnetic (FM) and different antiferromagnetic (AFM) states \cite%
{Kuebler83}. The parameters of the Heisenberg Hamiltonian were then fitted
to reproduce the course of the calculated energies but such technique
usually allows to extract only the parameters of the first and second
neighbors, and the interactions between Mn and all other atoms are neglected
for reasons of simplicity.

In this contribution, we derive the magnetic exchange parameters of a number
of Heusler alloys from first principles and then analyze the magnetic
coupling dependence on electronic structure variations induced by atomic
substitutions or volume variations. The paper is organized as follows: In
the next section \ref{sec:struc} we describe the crystal structure under
investigation and the computational method chosen. Section \ref%
{sec:grand-theory} is devoted to the parameter-free derivation of exchange
parameters by theoretical approaches within density-functional theory.
Section \ref{sec:results} contains our results for the electronic structures
and the magnetic interaction parameters of the Co$_{2}$MnZ, Rh$_{2}$MnZ and X%
$_{2}$MnSn families of alloys. Finally, we summarize our results in section %
\ref{sec:conclusion}.

\section{\label{sec:struc}Crystal structure and computational details}

\begin{figure}[th]
\includegraphics[scale=0.80,angle=0,origin=lb]{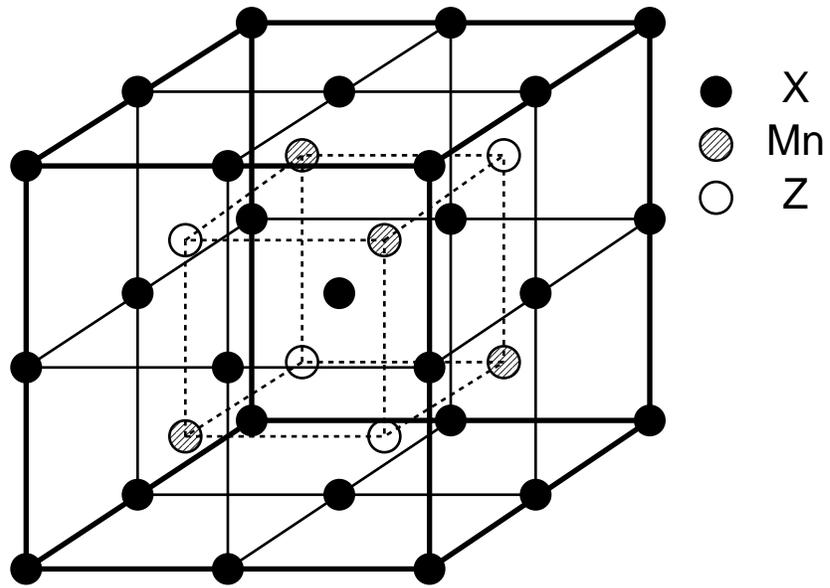}
\caption{The L2$_{1}$ structure type composed of four interpenetrating fcc
lattices.}
\label{fig:heusler_structure}
\end{figure}

\noindent The Heusler alloys represent a class of ternary intermetallic
compounds of general formula X$_{2}$YZ in which X is a transition metal, Z
is a metal of main groups III--V, and Y is a magnetically active transition
metal such as manganese. The Heusler alloys adopt ordered L2$_1$ structures,
given in Fig.\ \ref{fig:heusler_structure}, which may be understood as being
the result of four interpenetrating face-centered cubic (fcc) lattices.
According to the L2$_1$ structure jargon, the X atoms occupy the A and C
sites, the Y atoms are on the B sites and the Z atoms are found on the D
sites. Thus, sites A, B, C and D correspond to the positions (0,0,0), ($%
\frac{1}{4}$,$\frac{1}{4}$,$\frac{1}{4}$), ($\frac{1}{2}$,$\frac{1}{2}$,$%
\frac{1}{2}$) and ($\frac{3}{4}$,$\frac{3}{4}$,$\frac{3}{4}$) within the fcc
supercell \cite{69W1}.

The uniqueness of the Heusler alloys is due to the fact that they exhibit
cooperative magnetic phenomena --- especially ferromagnetism --- in the
desired temperature range although no constituent of their archetype, Cu$_2$%
MnAl, exhibits such properties in the elemental state. Even simpler than Cu$%
_2$MnAl is the phase MnAl which also displays strong ferromagnetic behavior
and is of technological interest because of an enhanced magnetic anisotropy.
The tetragonal MnAl ground state results from two subsequent (electronic and
structural) distortions away from a cubic structure \cite{mnal}. The group
of cubic Heusler alloys considered in this paper, however, all contain Mn
atoms as the Y atoms, and all phases exhibit ferromagnetic order. Their
lattice parameters, magnetic saturation moments and also experimental Curie
temperatures are shown in Tab.\ \ref{tab:tab_heusler}.

For the band-structure calculation we used the TB-LMTO-ASA method \cite%
{ole,TBLMTO}, including combined corrections, and took the experimental
values of the lattice parameters. The local-density approximation (LDA)
according to the Vosko--Wilk--Nusair exchange-correlation functional was
used \cite{vosko}. The summation over the entire Brillouin zone (BZ) was
performed with a total of 195 $\boldsymbol{k}$ points in the irreducible
part of the BZ.

\section{\label{sec:grand-theory}The calculation of exchange-interactions
parameters in density-functional theory}

\noindent The exchange coupling parameter $J_{ij}$ between two centers $i$
and $j$ being part of a magnetic material is usually defined in the
following standard procedure of the so-called rigid spin approximation
(RSA), 
\begin{equation}
J_{ij} = \boldsymbol{m}_{i}\frac{\partial ^{2}E}{\partial \boldsymbol{m}%
_{i}\partial \boldsymbol{m}_{j}}\boldsymbol{m}_{j} = \boldsymbol{m}_{i}\left[
\chi \right] _{ij}^{-1} \boldsymbol{m}_{j},  \label{e1}
\end{equation}
where $E$ is the total energy of the system, $\boldsymbol{m}_{i}$ is the
magnetic moment on site $i$, and $\chi _{ij}$ is a magnetic susceptibility.

In the above eq.\ \ref{e1}, the entry $\chi$ can be considered the adiabatic
(static) limit of the transversal part of the spin-dynamical susceptibility 
\begin{equation}
\chi (\boldsymbol{q},w)={\displaystyle \sum} \frac{\varphi _{\nu }^{\uparrow
}\left( \boldsymbol{k} \right) \varphi _{\nu ^{\prime }}^{\ast \uparrow }
\left( \boldsymbol{k}+\boldsymbol{q}\right) \varphi _{\nu }^{\downarrow
}\left( \boldsymbol{k}\right) \varphi _{\nu ^{\prime }}^{\ast \downarrow
}\left( \boldsymbol{k}+\boldsymbol{q}\right) } {\varepsilon _{\nu
}^{\uparrow }\left( \boldsymbol{k}\right) -\varepsilon _{\nu ^{\prime
}}^{\downarrow }\left( \boldsymbol{k}+\boldsymbol{q}\right) -\omega +i0},
\label{e2}
\end{equation}

where $\varphi _{\nu }^{\uparrow }\left( \boldsymbol{k}\right) $ and $%
\varepsilon _{\nu }^{\uparrow }\left( \boldsymbol{k}\right) $ are
eigenfunctions and eigenvalues of band structure problem, and the arrows
designate the spin direction. The so-called static limit $\left( \omega
\longrightarrow 0\right)$ can be justified if 
\begin{equation}
\varepsilon _{\nu }^{\uparrow }\left( \boldsymbol{k}\right) -\varepsilon
_{\nu ^{\prime }}^{\downarrow }\left( \boldsymbol{k}+\boldsymbol{q}\right) =%
\mathit{I}\text{ }m>>\omega.  \label{e3}
\end{equation}

This condition (eq.\ \ref{e3}) determines a whole range of spin-wave
frequencies for which one may use the adiabatic approximation and proceed
with the well-known Heisenberg model expression for the spin wave spectrum 
\begin{equation}
\omega \left( \boldsymbol{q}\right) =m\left( J(\boldsymbol{q})-J(0)\right)
=m\left( \chi ^{-1}\left( \boldsymbol{q}\right) -\chi ^{-1}\left( 0\right)
\right) .  \label{e4}
\end{equation}

Below we will estimate the validity of the above criterion (eq.\ \ref{e3})
for several compounds studied in this paper. Whenever eq.\ \ref{e3} is
satisfied in localized systems, however, eq.\ \ref{e1} can be further
simplified and a long-wave approximation (LWA, with essentially similar
smallness criteria as in the RSA) can be used to obtain the following
expression, 
\begin{equation}
J_{ij}^{\mathrm{lw}}=\boldsymbol{m}_{i}\left[ \chi \right] _{ij}^{-1}%
\boldsymbol{m}_{j}\approx \boldsymbol{m}_{i}\chi _{i}^{-1}\chi _{ij}\chi
_{j}^{-1}\boldsymbol{m}_{j}\approx \boldsymbol{m}_{i}I_{i}\chi _{ij}I_{j}%
\boldsymbol{m}_{j},  \label{e5}
\end{equation}%
where $\chi _{i}^{-1}$ is an on-site element of the inverted spin
susceptibility. Due to this similarity one cannot use the static
approximation (eq.\ \ref{e4} or any type of Heisenberg model) for large $q$
vectors. The model will be correct for large $q$ (or small distances in real
space) only if \emph{both} long-wave and adiabatic approximations are
removed simultaneously.

The currently most practical expression for the exchange coupling is based
on the Green function or multiple-scattering formalism. In this technique,
an analogue of eq.\ \ref{e1} can be derived \cite{me1} which reads 
\begin{eqnarray}
J(\boldsymbol{q}) &=& \frac{1}{N}{\displaystyle \sum} J_{ij}e^{i\boldsymbol{%
qR}_{ij}}=\frac{1}{N}{\displaystyle \sum} \frac{1}{\pi }\int\limits^{E_{F}}d%
\varepsilon \mathrm{ImTr}\left\{ \Delta _{i}\left[ T^{\uparrow
}T^{\downarrow }\right] _{ij}^{-1}\Delta _{j}\right\} e^{i\boldsymbol{qR}%
_{ij}}  \notag \\
&=& \frac{1}{\pi }\int\limits^{E_{F}}d\varepsilon \mathrm{ImTr}\left\{
\Delta _{i}\left[ \int d\boldsymbol{k}T^{\uparrow }\left( \boldsymbol{k}%
\right) T^{\downarrow }\left( \boldsymbol{k}+\boldsymbol{q}\right) \right]
^{-1}\Delta _{i}\right\}.  \label{e6}
\end{eqnarray}

where $\Delta _{i}=T_{ii}^{\uparrow }-T_{ii}^{\downarrow }$ and 
\begin{equation}
\text{ \ \ \ }T_{ij}^{\sigma }=\frac{1}{\Omega _{BZ}}\int d\boldsymbol{k}%
T^{\sigma }\left( \boldsymbol{q}\right) e^{i\boldsymbol{qR}_{ij}}=\frac{1}{%
\Omega _{BZ}}\int d\boldsymbol{k}\left( p^{\sigma }\left( \varepsilon
\right) -S\left( \boldsymbol{q}\right) \right) ^{-1}e^{i\boldsymbol{qR}%
_{ij}}.\text{ }  \label{e7}
\end{equation}
Here, $T^{\sigma }$ is a scattering path operator, $p^{\sigma }\left(
\varepsilon \right) $ is an atomic-potential scattering matrix and $S$ is a
matrix of structure constants. The corresponding long-wave limit was
obtained in Ref.\ \cite{lixt} and reads 
\begin{equation}
J^{\mathrm{lw}}(\boldsymbol{q})=\frac{1}{\pi \Omega _{BZ}}%
\int\limits^{E_{F}}\int d\boldsymbol{k}d\varepsilon \mathrm{ImTr}\left\{
p_{i}T^{\uparrow }(\boldsymbol{k})T^{\downarrow }\left( (\boldsymbol{k}+%
\boldsymbol{q})\right) p_{i}\right\},  \label{e8}
\end{equation}
where $p=p^{\uparrow }-p^{\downarrow }$. In real space, the zero-moment of
exchange interactions can be calculated accordingly to the sum rule 
\begin{equation}
J_{i}^{\mathrm{lw}}=\underset{j\neq i}{\displaystyle \sum}J_{ij}^{\mathrm{lw}%
}=\frac{1}{\pi }\int\limits^{E_{F}}d\varepsilon \mathrm{ImTr}[p_{i}\Delta
_{i}+p_{i}T_{i}^{\uparrow }T_{i}^{\downarrow }p_{i}].  \label{e9}
\end{equation}
The linearization of the multiple-scattering expression leads to the LMTO
Green function formalism where in a two-center approximation the $p^{\sigma
} $ matrix can be replaced by its linearized analogue 
\begin{equation}
p^{\sigma }\left( \varepsilon \right) =\frac{C^{\sigma }-\varepsilon }{%
\Delta ^{\sigma }},\text{ \ \ }  \label{e10}
\end{equation}
with spin-dependent LMTO potential parameters, namely $C^{\sigma}$ as the
band center and $\Delta^{\sigma }$ as its width. One can show \cite{Antropov}
that whenever $\Delta^{\uparrow }=\Delta ^{\downarrow }$, that is, equal
bandwidth for different spin projections, the Fourier transform of eq.\ \ref%
{e8} can be written in the form of eq.\ \ref{e5}. It is this limiting case
which allows us to separate the susceptibility and magnetic moment amplitude
contributions to the total exchange coupling. In the present paper, we will
mostly use eq.\ \ref{e8} because of its simplicity, but the applicability of
this approach will be checked and eq.\ \ref{e6} will be used if necessary.

Due to the presence of several magnetic atoms in a primitive cell, a
multi-atomic expression for the Curie temperature has to be used. In the
so-called mean field approximation (MFA), the Curie temperature $T_{\mathrm{C%
}}$ of the system with $N$ nonequivalent magnetic atoms is calculated as a
largest solution of the equation 
\begin{equation}
\mathop{\rm det}[T_{nm}-T\delta_{nm}]=0,  \label{e12}
\end{equation}
where $n$ and $m$ are the indices of the non-equivalent magnetic
sublattices, $T_{nm}=\frac{2}{3} J_{mn}^{0}$, and $J_{mn}^{0}$ is an
effective interaction of an atom from sublattice $n$ with all other atoms
from the sublattice $m$. In our case with two non-equivalent magnetic atoms
per cell, the expression for $T_{\mathrm{C}}$ is reduced to 
\begin{equation}
T_{\mathrm{C}}=\frac{1}{3}\{J_{\mathrm{Mn-Mn}}^{0}+J_{\mathrm{X-X}}^{0} + 
\sqrt{[J_{\mathrm{Mn-Mn}}^{0}-J_{\mathrm{X-X}}^{0}]^{2} + 4 \left( J_{%
\mathrm{Mn-X}}^{0}\right) ^{2}}\}.  \label{e13}
\end{equation}

\section{\label{sec:results}Results and discussion}

\subsection{\label{subsec:co2mnz}Co$_{2}$MnZ (Z = Ga, Si, Ge and Sn)
compounds}

\begin{figure}[th]
\includegraphics[scale=1.00,angle=0,origin=lb]{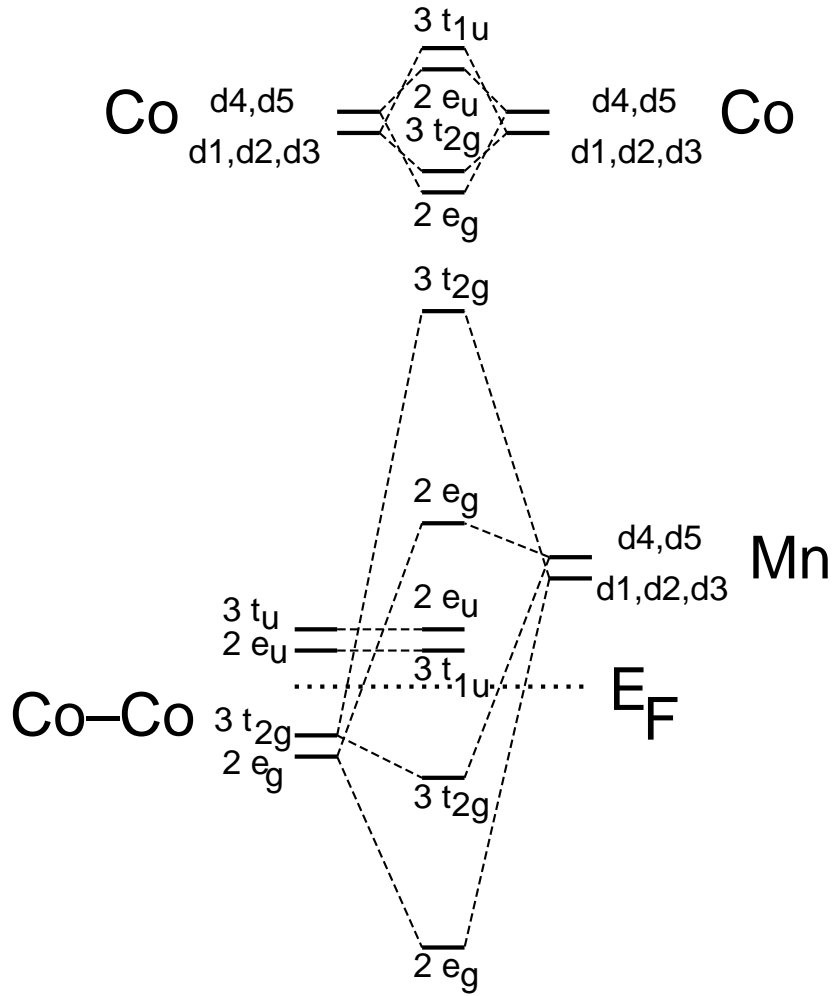}
\caption{Schematic hybridization between the minority spin orbitals of Co$%
_{2}$MnGa, first between two Co atoms (top), then between two Co atoms and a
neighboring Mn atom (bottom). The coefficients label the degeneracies of the
orbital sets (see notations in Ref.\ \protect\cite{Dederichs})}
\label{fig:fig2}
\end{figure}

\noindent To start with, we performed calculations of the electronic band
structures of four Co-based Heusler alloys with the generic formula Co$_{2}$%
MnZ (Z = Ga, Si, Ge and Sn). The results for the electronic spectra are in
good agreement with existing calculations of the electronic structures of
these compounds \cite%
{Kuebler83,Dederichs,Freeman,Nieminen,Ishida95_1,Ishida95_2}. To better
analyze the density-of-states (DOS) curves presented later, we first
schematically sketch the hybridizations \cite{chemistry-physics-jargon} of
the minority-spin orbitals between the Co and Mn atoms in Co$_2$MnGa, given
in Fig.\ \ref{fig:fig2}. It is justified to take the minority-spin orbitals
because, due to the exchange hole, these lie higher in energy and are
relatively diffuse such that they are much more involved in the chemical
bonding \cite{landrum-drons}. Their larger diffuseness also leads to the
finding that spin-polarized ground states show larger interatomic distances
despite of having a lower total energy \cite{drons-assp}. Compared to the
case of Co$_{2}$MnGe \cite{Dederichs}, the Fermi energy ($E_{\mathrm{F}}$)
in the Z = Ga case (see Fig.\ \ref{fig:fig2}) is placed below the Co--Co $t_{%
\mathrm{1u}}$ and $e_{\mathrm{u}}$ orbitals. For Co$_{2}$MnGe, the $t_{%
\mathrm{1u}}$ orbital is filled with one extra electron.

\begin{figure}[th]
\includegraphics[scale=0.80,angle=0,origin=lb]{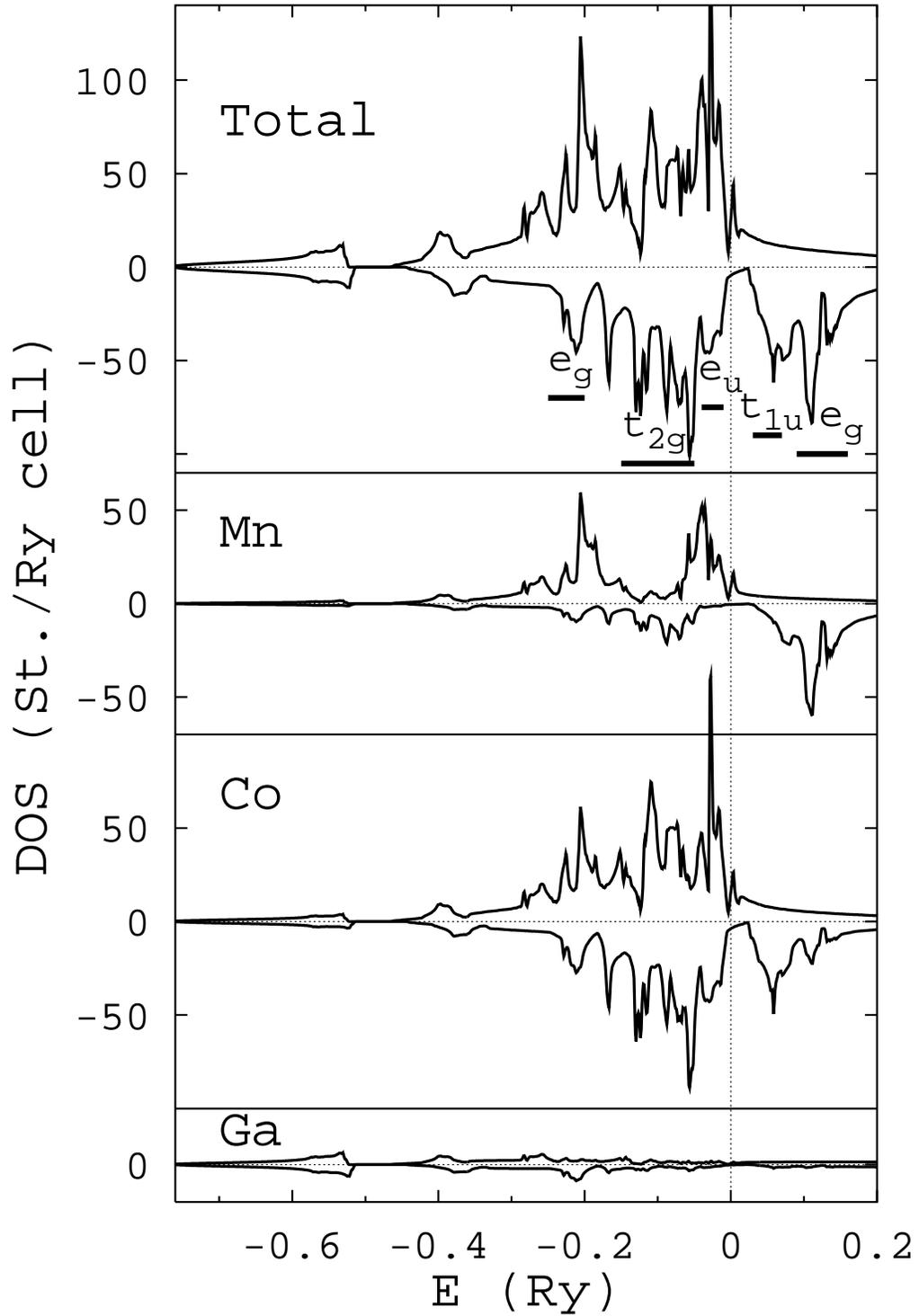}
\caption{Total and partial DOS for the compound Co$_{2}$MnGa. The character
of each peak belonging to the minority spin states has been indicated, and
the Fermi level is set to the energy zero.}
\label{fig:fig3}
\end{figure}

Fig.\ \ref{fig:fig3} presents the spin-polarized DOS of Co$_2$MnGa, and the
splittings between different symmetry states have been extracted at the zone
center $\Gamma$ for minority-spin states (see also discussion in Ref.\ \cite%
{Dederichs}). Note that this is an over-simplification because the symmetry
labels are not strictly valid over the entire reciprocal space. As expected,
the DOS around the Fermi level is heavily dominated by the $3d$ states of
the Mn and Co atoms, and the majority spin states are nearly fully occupied.
The DOS curves for the minority spins exhibit two peaks above the Fermi
level which are due to both Mn and Co $3d$ contributions. The difference in
the positions of these two peaks is directly determined by the difference in
intra-atomic exchange splitting between Mn and Co. The broad structure in
the lowest energy region between $-$0.8 and $-$0.55 Ry goes back to
(magnetically inactive) $4s$ and $4p$ states of Ga, and they are well
separated from the $3d$ states positioned in an energy region between $-$%
0.45 and 0.3 Ry. It is interesting to note that the Fermi level of Co$_2$%
MnGa is found at the DOS minimum of the minority states, but for Z = Si and
Ge it is positioned exactly in a gap, that is, these two latter compounds
exhibit 100\% spin polarization. This gap has previously been reported by
other authors and is formed due to the strong $3d$--$3d$ hybridization
(orbital mixing) between the Co and Mn atoms \cite{Ishida95_1, Ishida95_2}.

In such half-metallic compounds the total spin moment should ideally be an
integer number (see discussion in Ref.\ \cite{Dederichs}). Our results for
the Co$_{2}$MnZ group, presented in Tab.\ \ref{tab:tab_heusler}, are very
close to that, and there is only a slight deviation from integer numbers
reproducing so-called Slater--Pauling behavior: here, the total moment
equals $\mu _{\mathrm{tot}} = N -24$ where $N$ is the total number of
valence electrons in the unit cell. In accord with the DOS observation in
Fig.\ \ref{fig:fig3}, the Z ($sp$-type) atoms in Co$_2$MnZ have negligible
moments. The minority-spin states of the Mn atoms are nearly empty (see also
Fig.\ \ref{fig:fig3}), and the values of the local Mn spin moments arrive at
ca.\ 3 $\mu _{\mathrm{B}}$. The Co atoms do have significant spin moments,
about 0.7 $\mu_{\mathrm{B}}$ in Co$_2$MnGa and about 1.0 $\mu _{\mathrm{B}}$
in the remaining compounds of this family. Clearly, and also most
importantly, the exchange interactions between the Co and Mn atoms cannot be
neglected \emph{a priori}.

Accordingly, the calculated values of the partial contributions $J_{mn}^{0}$
to the effective exchange parameter $J_{n}^{0}$ are presented in Tab.\ \ref%
{tab:J0}. As has been alluded to already, the interaction between Mn and Co
atoms gives a \emph{leading} contribution to the total effective coupling,
thereby questioning the assumption used in earlier work \cite{Kuebler83} in
that only Mn--Mn interactions were taken into consideration; in terms of $3d$%
--$3d$ orbital overlap, this leading contribution is not at all surprising.
On the other side, the Co--Co interaction is negative ($-0.36$ mRy) in Co$%
_{2}$MnGa and thereby demonstrates the tendency for AFM ordering in the Co
sublattice. This negative value, however, is compensated by the larger \emph{%
positive} interaction between Co and Mn ($J_{\mathrm{Co-Mn}}^{0}\approx
2\times 2.9$ mRy) such that the \emph{effective} $J_{0}$ of Co remains large
and positive. The Mn--Mn contribution to $J_{\mathrm{Mn}}^{0}$ is on the
order of only 1 mRy, this is nearly five times smaller than the Co--Mn
interaction.

For completeness, we mention that pure $\alpha $-Mn exhibits an AFM ordering
at low temperature ($J_{0}<0$), and the small positive exchange parameters
by the nearest-neighbor Mn--Mn pairs in Mn-based Heusler compounds
correspond to likewise positive and small second-neighbor pairs exchange
parameters in pure $\alpha $-Mn. The theoretical values for the Curie
temperatures $T_{\mathrm{C}}$ obtained by the MFA are also included in Tab.\ %
\ref{tab:tab_heusler}. For Co$_{2}$MnGa, $T_{\mathrm{C}}$ arrives at 635 K
and underestimates the experimental value (694 K) by about 10\%. Taking into
account the fact that the MFA usually \emph{overestimates} Curie
temperatures, this must be considered an even larger disagreement with
experiment.

To check the nature of this disagreement, we performed a calculation of $J$\
beyond the LWA using eq.\ \ref{e7}. In Tab.\ \ref{tab:tab_heusler} we also
show the corresponding results obtained using this approach. Our
calculations reveal that the effective coupling between the Mn atoms is
practically unchanged so that the LWA is perfectly suitable for the
descripton of this coupling. However, all other couplings are affected much
more strongly by this approximation. For instance, the Co--Co interactions
are modified nearly by a factor of two while Co--Mn interactions are
increased by 25--35\% overall. Correspondingly, the estimated critical
temperatures of magnetic phase transition for this group of alloys are
increased by 10--20 \%, and they are larger than the experimentally observed
quantities.

It follows from the results presented in Tab.\ \ref{tab:J0} that the
substitution of the main-group III element Ga by a main-group IV element
such as Si, Ge or Sn leads to a significant increase of both $J_{\mathrm{%
Co-Mn}}^{0}$ and $J_{\mathrm{Mn-Mn}}^{0}$ values. Fortunately, the
implications for the varying electronic structure introduced by such a
substitution can be well described within the rigid-band approximation (RBA)
(see Ref.\ \cite{Dederichs,Ishida95_2}). For illustration, Fig.\ \ref%
{fig:fig4} shows a comparison between the electronic structures of Co$_2$%
MnGa and Co$_2$MnGe in the energy region $\pm 0.1$ Ry around the Fermi
level. The zero energy in the lower DOS curve corresponds to the Fermi level
of Co$_2$MnGe with 29 valence electrons; for Co$_2$MnGa with 28 valence
electrons, the Fermi level is given by the solid vertical line in the upper
DOS. Obviously, the DOS shapes for these two compounds are very similar to
each other, a nice support for the reliability of the RBA. Thus, by simply
changing the total number of valence electrons one may reproduce the
substitution of Ga by Ge fairly well, qualitatively.

Nonetheless, the total moment calculated from the electronic structure of Co$%
_{2}$MnGa only by extending the DOS to 29 valence electrons is just 4.6 $\mu
_{\mathrm{B}}$, that is, 8\% smaller than the numerical result (5 $\mu _{%
\mathrm{B}}$) for Co$_{2}$MnGe, and we will soon focus on this (small)
discrepancy originating from differing interatomic distances. In the frame
of the RBA, the significant increase in the exchange parameters (Tab.\ \ref%
{tab:J0}) in going from Co$_{2}$MnGa to Co$_{2}$MnGe goes back to the shift
of the Fermi energy (band filling in Ref.\ \cite{Schilfgaarde}) which
corresponds to the one extra electron. This evolution of the effective
parameters $J_{\mathrm{Mn}}^{0}=J_{\mathrm{Mn-Mn}}^{0}+J_{\mathrm{Mn-Co}%
}^{0} $ and also $J_{\mathrm{Co}}^{0}=J_{\mathrm{Co-Co}}^{0}+J_{\mathrm{Mn-Co%
}}^{0}/2$ as a function of the Fermi level is also included in Fig.\ \ref%
{fig:fig4}. In fact, the $J_{0}$ values of Co$_{2}$MnGa at the Fermi level
of Co$_{2}$MnGe equal those of the real Co$_{2}$MnGe phase. Taking into
consideration the usual MFA overestimation of the Curie temperatures, our
calculation for Co$_{2}$MnGe gives an acceptable agreement (1115 K) with the
experimental value of 905 K.

The dependence of the electronic structures and magnetic properties of the Co%
$_{2}$MnZ alloys on the chemical nature of the isoelectronic Z atom has
already been discussed in Refs.\ \cite{Freeman,Dederichs,Ishida95_2,Ishida90}%
. We will focus on the density of states (DOS) (see Fig.\ \ref{dos_co2mnz})
of Co$_{2}$MnZ (Z = Si, Ge and Sn) which all display the same valence
electron concentration. Not too surprisingly, the DOSs are similar to the
preceding one of Co$_{2}$MnGa. However, all peaks below the Fermi level move
to higher energies with increasing lattice parameters because of enlarging
atomic radii. The latter results in a smaller overlap between the Mn $3d$
and Co $3d$ orbitals which, in turn, leads to a smaller dispersion of these
bands \cite{Freeman}, becoming more atomic-like. As a consequence, the DOS
peaks come closer to each other and their amplitudes grow (Fig.\ \ref%
{dos_co2mnz}).

Because the changes in peak positions with changing Z element is
proportional to the change in the lattice parameter, the replacement of Si
by Ge has smaller consequences than the replacement of Ge by Sn; in terms of
radii (and chemical behavior), Si and Ge are more similar to each other.
Thus, the movement in the DOS peaks is more distinct for Co$_2$MnSn. In
agreement with the results of full-potential calculations \cite{Freeman},
the Mn magnetic moment obtained in our TB-LMTO-ASA calculation slightly 
\emph{increases} in the Si $\rightarrow$ Ge $\rightarrow $ Sn series. On the
other hand, the Co magnetic moment is \emph{lowered} so that the total
magnetic moment is close to 5 $\mu_{\mathrm{B}}$ in all three cases (see
Tab.\ \ref{tab:tab_heusler}). The increase of the Mn magnetic moment is
consistent with the increase of the Mn--Mn contribution to the effective $%
J_{0}$ (third column in Tab.\ \ref{tab:J0}) in this series of compounds. The
increase, however, is compensated by lower values for the Co--Co and also
Co--Mn interactions. Thus, the calculated $T_{\mathrm{C}}$'s decrease along
the line Si $\rightarrow$ Ge $\rightarrow$ Sn (see Tab.\ \ref%
{tab:tab_heusler}), and this qualitative trend agrees with the tendency
observed experimentally.

To fully demonstrate the volume dependence of the exchange interactions (and
Curie temperatures, too), we also calculated the course of $J_{0}$ in Co$_2$%
MnSi solely as a function of its volume. That is to say, the structure of Co$%
_2$MnSi was artificially expanded to lattice parameters that would better
fit the compounds adopted by its higher homologues Ge and Sn; unfortunately,
this is impossible to realize experimentally. Fig.\ \ref{j0_vol} displays
the values of the $J_{0}$ parameters obtained from these calculations, and
the purely volume-derived exchange parameters are in semi-quantitative
agreement with those that go back to proper calculations of $J_{0}$ for the
real Co$_{2}$MnGe and Co$_{2}$MnSn systems with their correct lattice
parameters. It is just too obvious that the behavior of the Curie
temperature can therefore be explained by a simple volume effect when the Si
atom is substituted by Ge or Sn.

A detailed comparison of Tabs.\ \ref{tab:tab_J_ij_co2mnz} and \ref{tab:vec}
in terms of $J_{ij}$ makes it clear that the interactions are relatively
short ranged and do not exceed the, say, first four neighbors in each
sublattice. The main exchange parameter, $J_{1}$ of Co$_{1}$--Mn in Tab.\ %
\ref{tab:tab_J_ij_co2mnz}, corresponds to the nearest-neighbor Co--Mn
interaction. This particular entry of the table alone already gives about
70\% of the total contribution to $J_{0}$ between Co and Mn atoms and is
about ten times larger than the corresponding Co--Co and Mn--Mn
interactions; a remarkable result but, as has been said before, not too
surprising when considering the interatomic overlap. For comparison, Tab.\ %
\ref{tab:tab_J_ij_co2mnz} also contains the exchange parameters obtained in
earlier work \cite{Kuebler83} where the authors calculated Mn--Mn exchange
parameters from the total energy differences of the FM and AFM structures
but by ignoring the Co--Mn interactions.

Naturally, their approach had to result in significantly larger Mn--Mn
interactions in order to reproduce the FM/AFM energy differences because in
such an approximation all interactions (Mn--Mn and Mn--Co) are effectively 
\emph{mapped} into the Mn--Mn-type $J_{ij}$. Thus, one needs to compare an
effective parameter for the Mn atom, namely $J_{\mathrm{Mn}}^{0}=J_{\mathrm{%
Mn-Mn}}^{0}+J_{\mathrm{Mn-Co}}^{0}$ from Tab.\ \ref{tab:J0} with $J_{\mathrm{%
Mn}}^{0}=12\times J_{1}+6\times J_{2}$ from Ref.\ \cite{Kuebler83}. The
obtained values are 10.9 mRy and 8.4 mRy correspondingly. This similarity
between the results of the very different models suggests relatively
localized magnetic character in this system.

\subsection{ \label{subsec:rh2mnz}Rh$_{2}$MnZ (Z = Ge, Sn and Pb) compounds}

\noindent Independent full-potential calculations of the electronic
structure of this group have been published recently \cite{Dederichs} and
our results are in agreement with them. To ease the understanding of the new
chemical system, we compare the densities-of-states of Rh$_2$MnSn with the
preceding one of Co$_2$MnSn, and both are included in Fig.\ \ref%
{dos_corh2mnsn}. Generally, the gap in the minority-spin states of the Co$_2$%
MnZ phases can also be observed for the Rh$_{2}$MnZ phases but this gap
apparently becomes broader and the Fermi level is no longer found in the
gap. Consequently, the total magnetic moment can no longer be an integer
number for this group of intermetallic compounds, and the entries of Tab.\ %
\ref{tab:tab_heusler} impressively support this statement.

Another important difference is given by the smaller width and also
polarization of the rhodium $3d$ states relative to those of cobalt. In
chemical terms, this notable difference between the $3d$ and $4d$ (and also $%
5d$) elements is easily explained by differences in spatial shielding, with
interesting similarities to main-group chemistry \cite{landrum-drons}. In
any case, the magnetic moment of the Rh atoms is only about half the size of
those of the Co atoms, namely ca.\ 0.45 $\mu_{\mathrm{B}}$ compared to ca.\
1 $\mu_{\mathrm{B}}$).

In contrast to the Co-based system, the Mn moments in this group are larger
by about 0.6 $\mu_{\mathrm{B}}$. Such a change has been explained \cite%
{Dederichs} by a smaller hybridization between the Rh and Mn atoms than
between the Co and Mn atoms. Alternatively, a chemical interpretation would
focus on an effectively over-sized Mn atom because of the strongly widened
lattice due to the large Rh atoms. Thus, the majority/minority spin
splitting for Mn is strongly favored, and the intra-atomic exchange
splitting will be mirrored by extraordinarily diffuse minority-spin orbitals
for Mn. The same effect takes place in FePd$_3$ where Fe acquires a very
large moment because of being too spacious \cite{landrum-drons}. Unlike the
results given in ref. \cite{Dederichs}, however, the total magnetic moments
of our calculations do not monotonically increase in the row Ge $\rightarrow$
Sn $\rightarrow$ Pb, but this effect is probably related to the atomic
spheres approximation used by us.

We will now analyze the results for the exchange coupling using eq.\ \ref{e5}%
. The values of the exchange splittings $m_{i}I_{i}$ for the Rh atoms are
about three times smaller than for the Co atoms so that Rh--Rh and Rh--Mn
exchange parameters are about ten and three times lower if compared to the
Co--Co and Co--Mn pairs (see Tab.\ \ref{tab:J0}) only because of this
splitting renormalization. Such a simple explanation, however, is not
applicable for the Mn--Mn interactions where the corresponding
susceptibility has also changed. In the Mn sublattice, the interactions are
decreased in magnitude by about 1 mRy upon substitution of Co by Rh despited
the increase of the Mn magnetic moments. Such a decrease for the Mn--Mn
exchange parameters reflects a general AFM tendency for a nearly half-filled 
$d$ band and an FM tendency for a nearly empty or filled $d$ band; this has
been discussed before \cite{AKL}. As can be seen from Fig.\ \ref%
{dos_corh2mnsn}, the manganese $d$ states in the Rh-based compounds are
nearly half filled while in Co-based compounds these Mn-centered states have
been filled by approx.\ 0.6 electrons despite Co/Rh being isoelectronic.

When it comes to the volume dependence of the magnetic properties of the
Rh-based compounds, we reiterate the course found for the Co$_{2}$MnZ (Z =
Ge, Sn, Pb) group (Fig.\ \ref{dos_co2mnz}). One also expects a decrease of
the Curie temperatures with increasing volume, and this is what the
experimental $T_{\mathrm{C}}$ values reflect in the row Ge $\rightarrow $ Sn 
$\rightarrow $ Pb (see Tab.\ \ref{tab:tab_heusler}). Unfortunately, this
trend is somewhat obscured in the theoretical data. The calculated Curie
temperatures in the LWA for Rh$_{2}$MnSn (435 K) and Rh$_{2}$MnPb (423 K)
are not too far away from the experimental ones, 412 and 338 K respectively.
For Rh$_{2}$MnGe, however, we underestimate $T_{\mathrm{C}}$ (410 K)
compared to an experimental 450 K. The usage of eq.\ \ref{e6} leads to the
significant modification of Rh--Rh coupling (factor of 2--3) and a 25--30\%
increase of the Rh--Mn coupling, \emph{i.e.}, the MFA produces significantly
larger numbers for $T_{\mathrm{C}}.$ However, all relative trends remain
similar to the exchange coupling in the LWA.

The energy dependence of $J$ in the Rh$_2$MnZ compounds with Z = Ge, Sn, Pb,
depicted in Fig.\ \ref{j0_rh2mnge}, looks different from the one discussed
before in the Co$_2$MnZ group. First, the amplitude of $J(E)$ is smaller
and, second, the maximum of the curve is a broad plateau. The last finding
means that an increase of the electron concentration will \emph{not} lead to
a significant change for the exchange parameters. An alternative decrease of
the electron concentration, however, leads to negative $J$ values such that
an FM state is no longer stable. For instance, the substitution of Ge by Al
shifts the Fermi level down by 0.04 Ry (vertical line in Fig.\ \ref%
{j0_rh2mnge}) and leads to a significant decrease of $J$. This
interpretation is supported by the experimental AFM ordering that was
observed for Rh$_2$MnAl \cite{Masumoto72}.

Closing this section, we'd like to mention that Rh$_{2}$MnZ compounds are
traditionally discussed as systems with fully localized magnetic moments, in
contrast to Co$_{2}$MnZ-type compounds where the Co magnetic moment can
obviously not be neglected. The results for the effective $J$ values in
Tab.\ \ref{tab:J0} and the pair-magnetic exchange values $J_{ij}$ in Tab.\ %
\ref{tab:tab_J_ij_rh2mnz} clearly evidence that Rh--Mn interactions are even
larger than Mn--Mn interactions. A similar behavior is known from Fe/Pd
alloys where the Fe atom magnetically polarizes the $4d$ metal upon strong
Fe--Pd chemical bonding \cite{landrum-drons}. In the present case, the
Rh--Mn exchange parameters are mostly determined by the first-neighbor $%
J_{1} $ interaction. Mn--Mn interactions show a significantly longer range
with the main contributions coming from large and positive $J_{1}$, $J_{2}$
and negative $J_{6}$.

\subsection{\label{subsec:x2mnsn}X$_2$MnSn (X = Ni, Cu and Pd) compounds}

\noindent In this section, we will analyze the change of the magnetic
properties of the Heusler alloys upon atomic substitution by the X
component, the non-Mn $d$ metal. For the compounds with X = Ni, Cu and Pd,
the electronic structures have been studied in Ref.\ \cite{Dederichs,
Kuebler83}. Similar to the preceding Rh$_2$MnZ group, the Fermi level is no
longer in the minority-spin DOS gap and the total moment is not an integer
number. The substitution of Co by Rh or Ni leads to a significant decrease
of the $d$-metal polarization and, also, to a nearly complete filling of
their minority-spin states. The magnetic moments of the X atoms is thereby
reduced from 1 $\mu_{\mathrm{B}}$ (Co) to ca.\ 0.5 $\mu_{\mathrm{B}}$ (Rh)
and, finally, to about 0.2 $\mu_{\mathrm{B}}$ (Ni), given in Tab.\ \ref%
{tab:tab_heusler}. This reduction is accompanied by an increase of the Mn
magnetic moment only during the first substitution. The limiting case is
given by the compounds with nonmagnetic Cu and Pd atoms.

Using the calculated magnetization values $m_{i}$, we can estimate the
reduction of the $J_{\mathrm{X-X}}^{0}$ and $J_{\mathrm{X-Mn}}^{0}$
parameters (see discussion above). The obtained parameters give qualitative
agreement with the directly calculated results, listed in Tab.\ \ref{tab:J0}%
. However, this estimation can not reproduce the decrease of $J_{\mathrm{%
Mn-Mn}}^{0}$ for X = Rh and, on the other hand, the significant increase for
X = Cu. The authors of Ref.\ \cite{Kuebler83} assumed that the principal
role of the X atoms is to simply determine the size of the crystal lattice.
To check this assumption, we calculated $J_{0}$ for Ni$_2$MnSn but with a
lattice parameter that is characteristic for Cu$_2$MnSn. As a result, $J_{%
\mathrm{X-Mn}}^{0}=1.3$ mRy and $J_{\mathrm{Mn-Mn}}^{0}=2.3$ mRy differ
strongly from the correctly calculated $J_{0}$ of real Cu$_2$MnSn by 0.3 mRy
and 5.7 mRy respectively (see Tab.\ \ref{tab:J0}). Also, the modified
exchange parameters upon $d$-metal substitution is not reproduced by the RBA
which worked nicely for an $sp$-component substitution.

In order to analyze the problem in more detail, we show the course of $J(E)$
as a function of Co$_{2}$MnSn band filling in Fig.\ \ref{j0_x2mnsn}. The
vertical lines correspond to the Fermi levels where the total number of
valence electrons is equal to the corresponding compound of the X$_2$MnSn
family. While it is clear that $J$ continuously decreases upon Co $%
\rightarrow$ Ni (and also Cu) substitution, this lowering is \emph{%
underestimated}, and the effective $J_{\mathrm{Mn}}^{0}$ obtained from Fig.\ %
\ref{j0_x2mnsn} is close to 8 mRy but the properly calculated $J_{\mathrm{Mn}%
}^{0}$ is 5 mRy (see Tab.\ \ref{tab:J0}).

The predicted Curie temperatures obtained from the calculated parameters are
presented in Tab.\ \ref{tab:tab_heusler}. The correct tendency for the
calculated Curie temperatures has been mentioned above, except for Ni$_{2}$%
MnSn where the disagreement is within the accuracy of the method. The total
exchange parameter $J_{0}$ is mostly determined by the first X--Mn pair
interaction and has significant long-range contributions; at least six
interactions are important, see Tab.\ \ref{tab:tab_J_ij_x2mnsn}. We also
include the results obtained from total-energy calculations \cite{Kuebler83}
and from a fit to spin-wave dispersions \cite{Noda}. The exchange coupling
in the LWA produces somewhat smaller values for Mn--Mn interaction while
X--Mn interactions are underestimated by 50-60\% when compared with those
from the general definition (eq.\ \ref{e6}). All $T_{\mathrm{C}}$'s are
overestimated in this approach and we expect that any improvement of the MFA
will produce better aggreement with experiment. The calculated exchange
parameters can be used in any more sophisticated calculations of the
critical temperature.

As mentioned before, one can compare the Mn total exchange only. The $J_{ij}$
obtained in Ref.\ \cite{Kuebler83} are in good agreement with our results
for Pd$_{2}$MnSn (2.1 mRy and 2.5 mRy respectively), but for the Ni- and
Cu-based compounds, the authors obtained numbers which are two to three
times smaller than ours. The result obtained from spin-wave dispersions in Ni%
$_{2}$MnSn (3.3 mRy) is fairly close to our 4.4 mRy but for Pd$_{2}$MnSn,
however, the disagreement is significant (1.3 mRy versus 2.5 mRy).
Nonetheless, it must be mentioned that both calculations of exchange
parameters did \emph{not} include the Mn--X interactions which are important
especially in the Ni$_{2}$MnSn system. The results obtained from the energy
differences of FM and AFM ordered structures tend to give systematically
underestimated exchange parameters although these systems are considered as
localized-moment systems. The results from spin-wave analysis also
underestimate the exchange coupling. We therefore plan to consider the
spin-wave properties in future publications.

\section{\label{sec:conclusion}Conclusion}

\noindent The electronic structures and magnetic exchange interactions have
been calculated for a set of full-Heusler alloys with generic formula X$_{2}$%
MnZ where X is a transition-metal atom and Z is an $sp$ main-group element.
The alloy variations of the Curie temperatures calculated in the mean-field
approximation are in good agreement with experimental data. Our analysis
demonstrates that the $J_{ij}$ dependence on the Z atom may be described
within a rigid band approximation, having straightforward implications for
the influence of the atomic volume of Z, thereby allowing semi-quantitative
predictions. The substitution of an X element, however, poses a problem for
the rigid-band approximation although qualitative tendencies can be
identified; for obtaining quantitative results, a full calculation has to be
performed. The magnetic exchange parameters and also Curie temperatures
decrease along the row Cu $\rightarrow $ Ni $\rightarrow $ Rh $\rightarrow $
Pd, in agreement with the degree of $d$ localization for the transition
metal. The X--Mn interactions are very important for systems with sizable
magnetic moments on the transition metal (Co, Rh and Ni). The X--Mn
interactions are limited by first neighbors while Mn--Mn interactions are
quite long ranged.

This work was carried out, in part, at Ames Laboratory, which is operated
for the U.S. Department of Energy by Iowa State University under Contract
No. W-7405-82. This work was supported by the Director for Energy Research,
Office of Basic Energy Sciences of the U.S. Department of Energy. The
support by Deutsche Forschungsgemeinschaft (Grant No.\ DR 342/7-1) is also
gratefully acknowledged.

\clearpage

\begin{table}[tbp]
\caption{Experimental lattice parameters $a$, calculated partial and
experimental total magnetic moments $\protect\mu $, and calculated and
experimental Curie temperatures $T_{\mathrm{C}}$ for X$_{2}$MnZ compounds.
All experimental values have been taken from Ref.\ \protect\cite{Webster}}%
\begin{ruledtabular}
\begin{tabular}{lcccccccc}
compound & $a$ (a.u.) & \multicolumn{3}{c}{$\mu_{\rm calc}$ ($\mu_{\rm B}$)} & 
                                           $\mu_{\rm expt}$ ($\mu_{\rm B}$) &  
                        \multicolumn{3}{c}{$T_{\rm C}$ (K)} \\
\cline{3-9}
           &        & X & Mn & total &  total  & LWA  &  exact &expt\\
\hline
\hline
Co$_2$MnGa & 10.904 & 0.73 & 2.78 & 4.13 & 4.05 & 635  & 880 & 694 \\ 
\hline 
Co$_2$MnSi & 10.685 & 1.01 & 3.08 & 5.00 & 5.07 & 1251 & 1563 & 985 \\ 
\hline 
Co$_2$MnGe & 10.853 & 0.97 & 3.14 & 5.00 & 5.11 & 1115& 1417 & 905 \\ 
\hline
Co$_2$MnSn & 11.338 & 0.95 & 3.24 & 5.04 & 5.08 & 1063 &1325 & 829 \\ 
\hline 
\hline
Rh$_2$MnGe & 11.325 & 0.42 & 3.67 & 4.49 & 4.62 & 410  & 549 & 450 \\ 
\hline 
Rh$_2$MnSn & 11.815 & 0.45 & 3.73 & 4.60 & 3.10 & 435  & 585 & 412 \\ 
\hline 
Rh$_2$MnPb & 11.966 & 0.45 & 3.69 & 4.58 & 4.12 & 423  &579 & 338 \\
\hline 
\hline 
Ni$_2$MnSn & 11.439 & 0.23 & 3.57 & 3.97 & 4.05 & 373  & 503 & 344 \\ 
\hline 
Cu$_2$MnSn & 11.665 & 0.04 & 3.79 & 3.81 & 4.11 & 602  & 680 & 530 \\
\hline 
Pd$_2$MnSn & 12.056 & 0.07 & 4.02 & 4.07 & 4.23 & 232  & 275 & 189 \\
\end{tabular}
\end{ruledtabular}
\label{tab:tab_heusler}
\end{table}

\clearpage

\begin{table}[tbp]
\caption{Sublattices contributions $J_{nm}^{0}$ (in mRy) to the effective
magnetic exchange parameters $J_{n}^{0}=\sum_{m}J_{nm}^{0}$ for the X$_{2}$%
MnZ group of compounds (both long wave approximation and exact adiabatic
results are shown).}
\label{tab:J0}%
\begin{ruledtabular}
\begin{tabular}{lrrrrrr}
compound   & \multicolumn{2}{c}{X--X}   & \multicolumn{2}{c}{X--Mn}  & 
             \multicolumn{2}{c}{Mn--Mn} \\
                        & $J^{\rm lw}$ & $J$ &  $J^{\rm lw}$ & $J$ & $J^{\rm lw}$& J \\
\hline
\hline
Co$_2$MnGa & $-0.36$ & $-0.21$ & 5.8 & 7.31& 0.81 & 0.89 \\
\hline
Co$_2$MnSi  &  1.57 & 2.6 & 10.2 & 11.4& 1.83 & 1.85 \\
\hline
Co$_2$MnGe &  1.12 & 2.3 & 8.92& 10.1 & 2.20 & 2.2 \\
\hline
Co$_2$MnSn &  0.55 & 0.94 & 8.66 & 9.9 & 2.24 & 2.3 \\
\hline
\hline
Rh$_2$MnGe &  0.06 & 0.2 & 3.17 & 4.0 & 1.28 & 1.31\\
\hline 
Rh$_2$MnSn &  0.11 & 0.25& 3.38 & 4.25 & 1.29 & 1.38 \\
\hline
Rh$_2$MnPb &  0.14 & 0.38& 3.24 & 4.15 & 1.32 & 1.35 \\
\hline
\hline
Ni$_2$MnSn & $-0.064$ & $-0.1$ & 1.92 & 2.8 & 2.52 & 2.63\\
\hline
Cu$_2$MnSn &  0.00 & 0.01& 0.26 & 0.40 & 5.71 & 5.9\\
\hline
Pd$_2$MnSn &  0.00 & 0.01& 0.29 & 0.41& 2.17 & 2.34\\
\end{tabular}
\end{ruledtabular}
\end{table}

\clearpage

\begin{table}[tbp]
\caption{Pair exchange interaction parameters $J_{ij}$ (in $\protect\mu $Ry)
in the long-wave approximation for the Co$_{2}$MnZ (Z = Ga, Si, Ge or Sn)
family and results from Ref.\ \protect\cite{Kuebler83}.}
\label{tab:tab_J_ij_co2mnz}%
\begin{ruledtabular}
\begin{tabular}{llrrrrrrrr}
compound & sublatt. &$J_1$&$J_2$&$J_3$&$J_4$&$J_5$&$J_6$&$J_7$&$J_8$ \\
\hline
\hline
Co$_2$MnGa &Co$_1$--Co$_1$&$-11$& 4&$-7$&$-7$& 2& 2& 1& 1\\ 
\cline{2-10}
           &Co$_1$--Co$_2$& 49& 4&$-76$&$-5$& 0&$-3$& 2& 1\\ 
\cline{2-10}
           & Co$_1$--Mn   & 557& 64&$-5$&$-3$&$-1$&$-3$& 0& 0\\ 
\cline{2-10}
           & Mn--Mn       & 36&$-2$& 4& 17& 2&$-3$& 8& 3\\ 
\hline
Co$_2$MnSi &Co$_1$--Co$_1$& 5& 59& 1& 1& 1&$-2$& 1& 1\\ 
\cline{2-10}
           &Co$_1$--Co$_2$& 165& 72&$-31$&$-6$& 10& 0& 2& 1\\ 
\cline{2-10}
           & Co$_1$--Mn   & 1106& 38& 12& 2& 4& 1& 0& 0\\ 
\cline{2-10}
           & Mn--Mn       & 130& 58&$-12$& 24& 0&$-8$& 0&$-2$\\ 
\hline
Co$_2$MnGe &Co$_1$--Co$_1$&$-11$& 55& 2& 2& 1&$-2$& 0& 0\\ 
\cline{2-10}
           &Co$_1$--Co$_2$& 136& 75&$-53$&$-5$& 10&$-1$& 2& 1\\ 
\cline{2-10}
           & Co$_1$--Mn   & 932& 41& 11& 3& 6& 1& 0& 0\\ 
\cline{2-10}
           & Mn--Mn       & 141& 60&$-5$& 23& 1&$-5$& 1&$-4$\\ 
\hline
Co$_2$MnSn &Co$_1$--Co$_1$&$-40$& 57& 4& 4& 2&$-5$& 0& 0\\ 
\cline{2-10}
           &Co$_1$--Co$_2$& 73& 87&$-56$&$-6$& 13&$-4$& 2& 4\\ 
\cline{2-10}
           & Co$_1$--Mn   & 907& 40& 7& 0& 9& 2& 4& 1\\ 
           & Co$_1$--Mn   & 907& 40& 7& 0& 9& 2& 4& 1\\ 
\cline{2-10}
           & Mn--Mn       & 126& 78& 5& 26&$-3$&$-13$& 0& 0\\ 
\cline{2-10}
           & Mn--Mn$^{[9]}$ & 630& 135& & & & & & \\ 
\end{tabular}
\end{ruledtabular}
\end{table}

\clearpage

\begin{table}[tbp]
\caption{The radius-vector $\boldsymbol{r}$, the distance from the central
atom $r$, and the number of equivalent nearest neighbors $n$ for the L2$_{1}$
type of structures.}
\label{tab:vec}%
\begin{ruledtabular}
\begin{tabular}{ccccccccccccc}
$J_i$&\multicolumn{3}{c}{X$_1$--X$_1$}&\multicolumn{3}{c}{X$_1$--X$_2$}&
      \multicolumn{3}{c}{X$_1$--Mn}&\multicolumn{3}{c}{Mn--Mn}\\
\cline{2-13}
     &$n$ & $r$ & $\boldsymbol r$ &$n$ & $r$ & $\boldsymbol r$ 
     &$n$ & $r$ & $\boldsymbol r$ &$n$ & $r$ & $\boldsymbol r$\\
\hline
\hline
$J_1$&12&0.707& $\bar{\frac{1}{2}} \bar{\frac{1}{2}} 0$ & 6&0.500&$ 0 \frac{1}{2} 0$ & 
       4&0.433& $\frac{1}{4} \frac{1}{4} \frac{1}{4}$ &12&0.707&$\bar{\frac{1}{2}} \bar{\frac{1}{2}}0$\\
$J_2$& 6&1.000& $0 0 \bar{1}$ & 4&0.866& $\bar{\frac{1}{2}} \bar{\frac{1}{2}} \bar{\frac{1}{2}}$ &
      12&0.829& $\bar{\frac{1}{4}} \frac{3}{4} \frac{1}{4}$ & 6&1.000& $0 0 \bar{1}$ \\
$J_3$&12&1.225& $\bar{\frac{1}{2}} \bar{\frac{1}{2}} \bar{1}$ & 4&0.866& $\bar{\frac{1}{2}} \bar{\frac{1}{2}}
                                                                                          \frac{1}{2}$ &
      12&1.090& $\frac{3}{4} \frac{1}{4} \frac{3}{4}$ &24&1.225& $\bar{\frac{1}{2}} \bar{\frac{1}{2}} \bar{1}$\\
$J_4$&12&1.225& $1 \bar{\frac{1}{2}} \bar{\frac{1}{2}}$ &24&1.118& $\bar{\frac{1}{2}} 0 \bar{1}$ &
      12&1.299& $\bar{\frac{5}{4}} \bar{\frac{1}{4}} \frac{1}{4}$ &12&1.414& $\bar{1} \bar{1} 0$ \\
$J_5$&12&1.414& $\bar{1} \bar{1} 0$ & 6&1.500& $0 0 \bar{\frac{3}{2}}$ &
       4&1.299& $\bar{\frac{3}{4}} \frac{3}{4} \frac{3}{4}$ &24&1.581& $0 \bar{\frac{1}{2}} \bar{\frac{3}{2}}$ \\
$J_6$&24&1.581& $0 \bar{\frac{1}{2}} \bar{\frac{3}{2}}$ &12&1.500& $\bar{1} \bar{\frac{1}{2}} \bar{1}$ &
      24&1.479& $\bar{\frac{5}{4}} \bar{\frac{3}{4}} \bar{\frac{1}{4}}$ & 8&1.732& $\bar{1} \bar{1} \bar{1}$\\
$J_7$& 4&1.732& $\bar{1} \bar{1} \bar{1}$ &12&1.500& $\bar{1} \frac{1}{2} \bar{1}$ &
      12&1.639& $\bar{\frac{3}{4}} \bar{\frac{5}{4}} \frac{3}{4}$ &48&1.871& $\frac{3}{2} 1 \frac{1}{2}$ \\
$J_8$& 4&1.732& $\bar{1} 1 \bar{1}$ &12&1.658& $\frac{1}{2} \bar{\frac{3}{2}} \frac{1}{2}$ &
      12&1.785& $\frac{1}{4} \bar{\frac{7}{4}} \frac{1}{4}$ & 6&2.000& $0 0 \bar{2}$ \\
\end{tabular}
\end{ruledtabular}
\end{table}

\clearpage

\begin{table}[tbp]
\caption{Pair magnetic exchange interactions $J_{ij}$ (in $\protect\mu $Ry)
in the long-wave approximation calculated for Rh$_{2}$MnZ (Z = Ge, Sn or
Pb). }
\label{tab:tab_J_ij_rh2mnz}%
\begin{ruledtabular}
\begin{tabular}{llrrrrrrrr}
compound & sublatt. &$J_1$&$J_2$&$J_3$&$J_4$&$J_5$&$J_6$&$J_7$&$J_8$ \\
\hline
\hline
Rh$_2$MnGe &Rh$_1$--Rh$_1$&$-9$& 13& 0& 0&$-1$& 0& 0& 0\\ 
\cline{2-10}
           &Rh$_1$--Rh$_2$& 17& 0&$-3$&$-2$& 6& 0& 0& 1\\ 
\cline{2-10}
           & Rh$_1$--Mn   & 312& 21& 2& 2& 2& 1& 0&$-1$\\ 
\cline{2-10}
           & Mn--Mn       & 87& 102& 3& 18&$-5$&$-27$&$-4$&$-9$\\ 
\hline
Rh$_2$MnSn &Rh$_1$--Rh$_1$ &$-8$& 15& 0& 0&$-2$& 0& 0& 0\\ 
\cline{2-10}
           &Rh$_1$--Rh$_2$& 20& 2&$-3$&$-2$& 8&$-1$& 0& 1\\ 
\cline{2-10}
           & Rh$_1$--Mn   & 340& 20& 3& 2& 2& 1& 1&$-1$\\ 
\cline{2-10}
           & Mn--Mn       & 61& 108& 6& 29&$-6$&$-27$&$-4$&$-3$\\ 
\hline
Rh$_2$MnPb &Rh$_1$--Rh$_1$&$-8$& 15& 0& 0&$-2$& 1&$-1$&$-1$\\ 
\cline{2-10}
           &Rh$_1$--Rh$_2$& 22& 2&$-5$&$-2$& 7&$-1$& 0& 1\\ 
\cline{2-10}
           & Rh$_1$--Mn   & 327& 17& 5& 2& 1& 1& 1&$-1$\\ 
\cline{2-10}
           & Mn--Mn       & 57& 96& 21& 33&$-8$&$-34$&$-9$&$-4$\\ 
\end{tabular}
\end{ruledtabular}
\end{table}

\clearpage

\begin{table}[tbp]
\caption{Pair magnetic exchange interactions $J_{ij}$ (in $\protect\mu $Ry)
in the long-wave approximation calculated for X$_{2}$MnSn (X = Ni, Cu or Pd)
and results from Refs.\ \protect\cite{Kuebler83,Noda}.}
\label{tab:tab_J_ij_x2mnsn}%
\begin{ruledtabular}
\begin{tabular}{llrrrrrrrr}
compound & sublatt. &$J_1$&$J_2$&$J_3$&$J_4$&$J_5$&$J_6$&$J_7$&$J_8$ \\
\hline
\hline
Ni$_2$MnSn & Ni$_1$--Mn   & 263&$-18$& 1& 4& 8& 1& 1& 2\\ 
\cline{2-10}
           & Mn--Mn       & 151& 116& 29&$-104$& 14&$-30$& 12&$-14$\\ 
\cline{2-10}
           & Mn--Mn$^{[9]}$ & 187&$-13$& & & & & & \\ 
\cline{2-10}
           & Mn--Mn$^{[16]}$ & 82& 105& 38& 37&$-6$& 17& 4& 2\\ 
\hline
Cu$_2$MnSn & Cu$_1$--Mn   & 30& 2& 0& 0&$-1$& 0& 0& 0\\ 
\cline{2-10}
           & Mn--Mn       & 491& 318&$-118$& 19&$-12$& 65& 9& 9\\ 
\cline{2-10}
           & Mn--Mn$^{[9]}$ & 88& 97& &  & & & & \\ 
\hline
Pd$_2$MnPb & Pd$_1$--Mn   & 40&$-3$& 0& 0& 2& 0& 0& 0\\ 
\cline{2-10}
           & Mn--Mn       & 65& 116& 51&$-78$& 20&$-64$& 16&$-5$\\ 
\cline{2-10}
           & Mn-Mn$^{[9]}$ & 187&$-20$& & & & & & \\ 
\cline{2-10}
           & Mn-Mn$^{[16]}$ & 64& 43& 21&$-44$& 14&$-19$& 4&$-6$\\ 
\end{tabular}
\end{ruledtabular}
\end{table}

\clearpage

\clearpage

\clearpage

\clearpage

\begin{figure}[th]
\includegraphics[scale=0.70,angle=0,origin=lb]{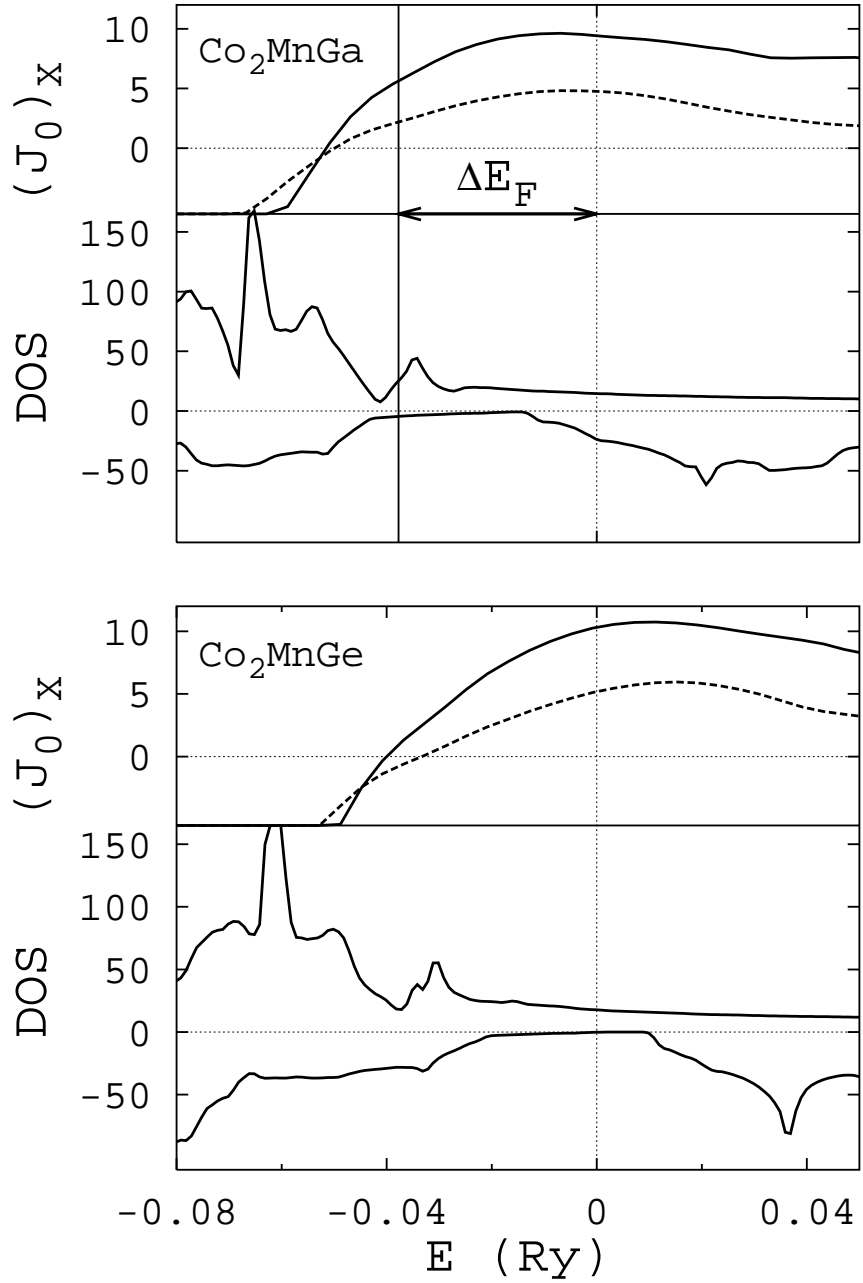}
\caption{Density of states and $J^{0}(E)$ for the Mn (solid line) and Co
(dashed line) atoms of Co$_{2}$MnGa (top) and Co$_{2}$MnGe (bottom). The
Fermi level is at zero energy for Co$_{2}$MnGe (29 valence electrons) and
shifted to the left for Co$_{2}$MnGa (28 valence electrons) by the
rigid-band shift $\Delta E_{\mathrm{F}}$.}
\label{fig:fig4}
\end{figure}

\clearpage

\begin{figure}[th]
\includegraphics[scale=0.60,angle=270,origin=lb]{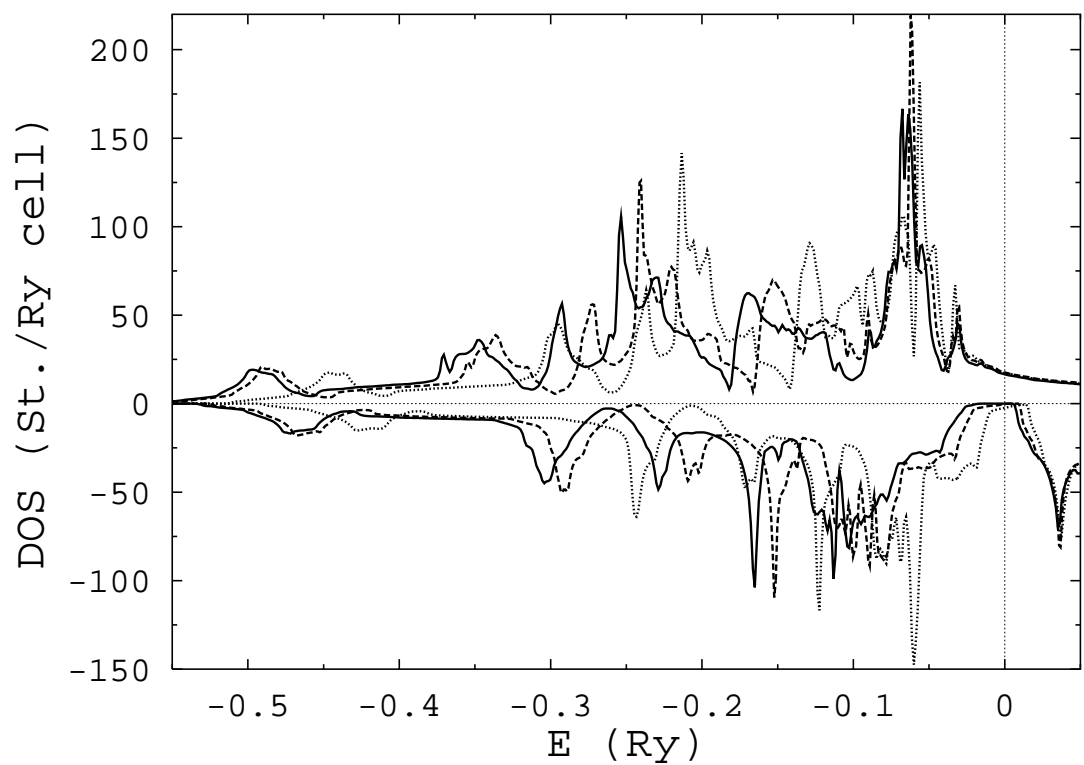}
\caption{Density of states of the compounds Co$_{2}$MnSi (solid line), Co$%
_{2}$MnGe (dashed line) and Co$_{2}$MnSn (dotted line). The Fermi level is
at the energy zero.}
\label{dos_co2mnz}
\end{figure}

\clearpage

\begin{figure}[th]
\includegraphics[scale=0.60,angle=270,origin=lb]{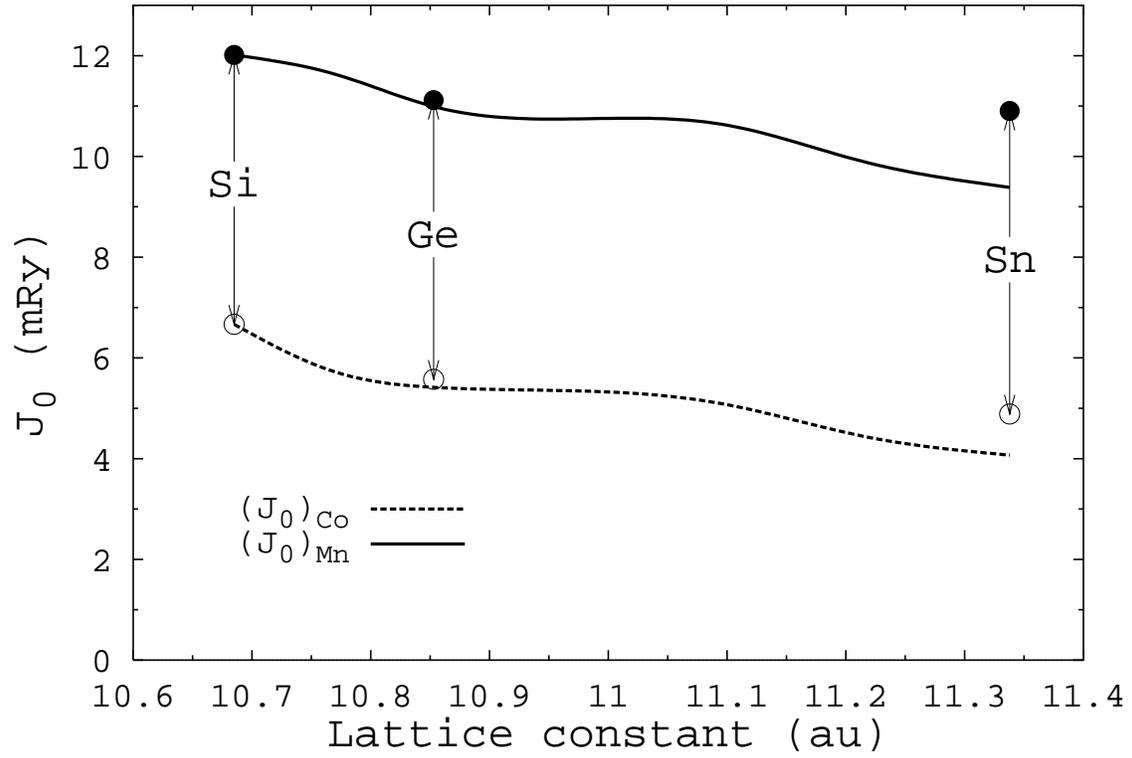}
\caption{Course of $J_{\mathrm{Mn}}^{0}$ (solid line) and $J_{\mathrm{Co}%
}^{0}$ (dashed line) as a function of the lattice parameter in Co$_{2}$MnSi.
Filled circles correspond to $J_{\mathrm{Mn}}^{0}$ and empty ones to $J_{%
\mathrm{Co}}^{0}$ for Co$_{2}$MnZ systems (Z = Si, Ge and Sn) calculated at
their experimental lattice parameters.}
\label{j0_vol}
\end{figure}

\clearpage

\begin{figure}[th]
\includegraphics[scale=0.60,angle=0,origin=lb]{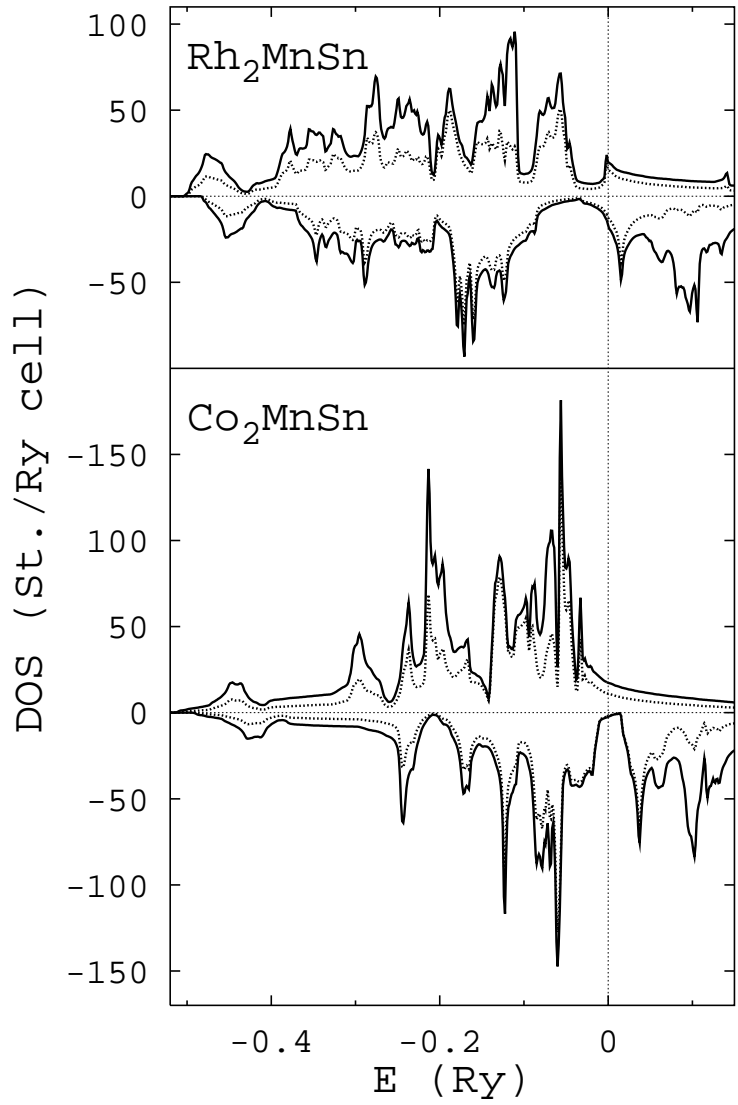}
\caption{Total (solid line) and partial densities of states (dashed line) of
Mn in Rh$_{2}$MnSn and Co$_{2}$MnSn. The Fermi level is at the energy zero.}
\label{dos_corh2mnsn}
\end{figure}

\clearpage

\begin{figure}[th]
\includegraphics[scale=0.60,angle=270,origin=lb]{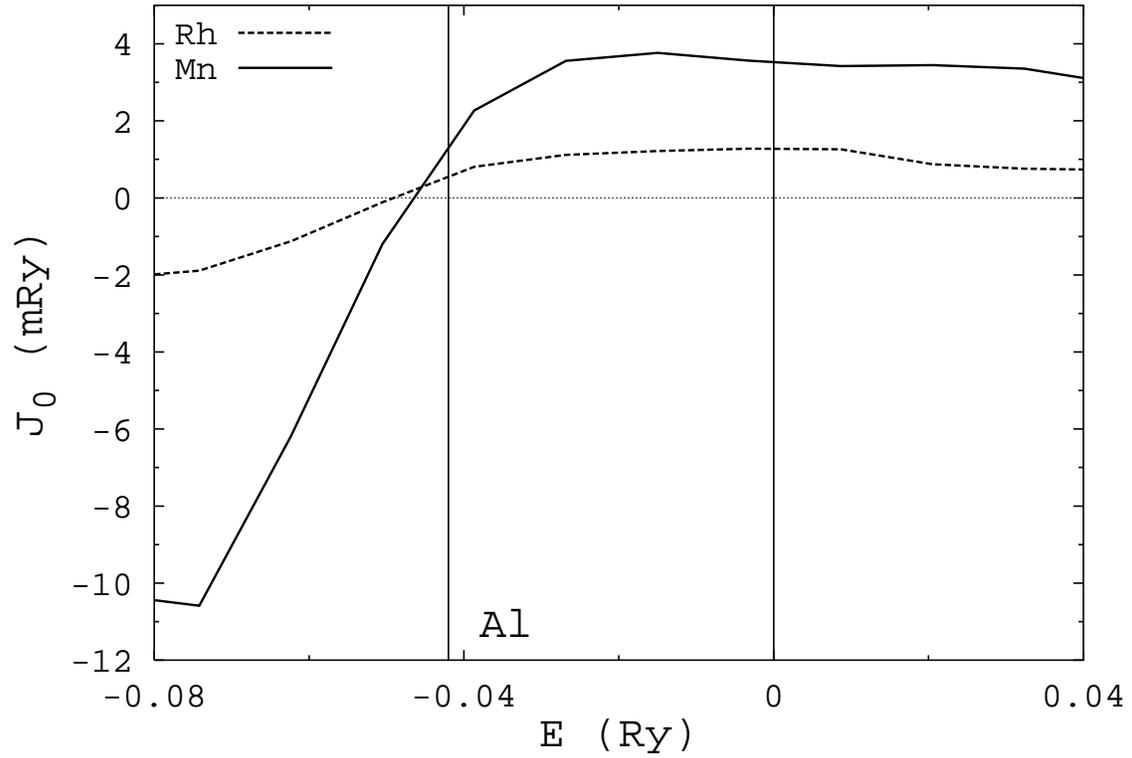}
\caption{The calculated effective exchange parameters $J_{\mathrm{Mn}}^{0}$
(solid) and $J_{\mathrm{Rh}}^{0}$ (dashed line) as a function of band
filling for Rh$_{2}$MnGe. Vertical lines corresponds to 28 (Rh$_{2}$MnAl)
and 29 (Rh$_{2}$MnGe) electrons per unit cell. The energy zero corresponds
to the Fermi level of Rh$_{2}$MnGe.}
\label{j0_rh2mnge}
\end{figure}

\clearpage

\begin{figure}[th]
\includegraphics[scale=0.60,angle=270,origin=lb]{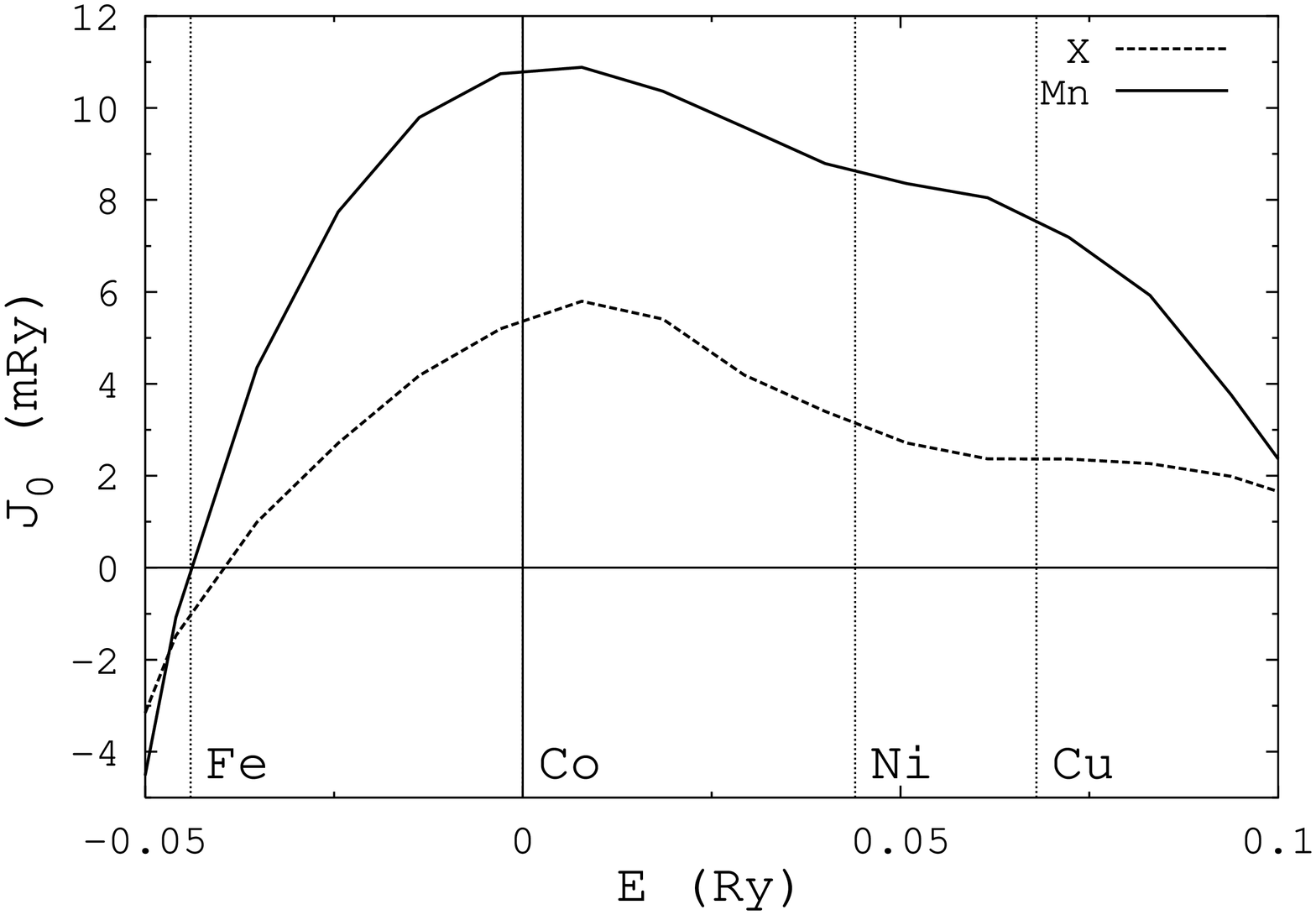}
\caption{The calculated effective parameters $J_{\mathrm{Mn}}^{0}$ (solid
line) and $J_{\mathrm{Co}}^{0}$ (dashed line) as a function of band filling
in Co$_{2}$MnSn. Vertical lines correspond to 27 (Fe$_{2}$MnSn), 29 (Co$_{2}$%
MnSn), 31 (Ni$_{2}$MnSn) and 33 (Cu$_{2}$MnSn) electrons per unit cell.}
\label{j0_x2mnsn}
\end{figure}


\begin{thebibliography}{99}
\bibitem{deGroot83} R. de Groot, P. van Engen, Phys. Rev. Lett. \textbf{50}
(1983) 2024.

\bibitem{Ishida98} S. Ishida, T. Masaki, S. Fujii, S. Asano, Physica B 
\textbf{245} (1998) 1.

\bibitem{Julliere75} M. Julliere, Phys. Lett. A \textbf{54} (1975) 225.

\bibitem{Moodera95} J. S. Moodera, L. R. Kinder, T. M. Wong, R. Meservey,
Phys. Rev. Lett. \textbf{74} (1995) 3273.

\bibitem{Dieny91} B. Dieny, V. S. Speriosu, S. S. P. Parkin, B. A. Gurney,
D. R. Wilhoit, D. Mauri, Phys. Rev. B \textbf{43} (1991) 1297.

\bibitem{heusler} O. Heusler, Ann. Phys. \textbf{19} (1934) 155.

\bibitem{Ishikawa} Y. Ishikawa, Physica B \textbf{91} (1977) 130.

\bibitem{Hamzic} A. Hamzi\'{c}, R. Asomoza, I.A. Campbell, J. Phys. F 
\textbf{11} (1981) 1441.

\bibitem{Kuebler83} J. K\"ubler, A. R. Williams, C. B. Sommers, Phys. Rev. B 
\textbf{28} (1983) 1745.

\bibitem{Dederichs} I. Galanakis, P. H. Dederichs, N. Papanikolaus, Phys.
Rev. B \textbf{66} (2002) 174429.

\bibitem{Freeman} S. Picozzi, A. Continenza, A. J. Freeman, Phys. Rev. B 
\textbf{66} (2002) 094421.

\bibitem{Nieminen} A. Ayuela, J. Enkovaara, K. Ullakko, R. M. Niemenen, J.
Phys.: Condens. Matter \textbf{11} (1999) 2017.

\bibitem{Ishida95_1} S. Fujii, S. Ishida, S. Asano, J. Phys. Soc. Jpn 
\textbf{64} (1995) 185.

\bibitem{Ishida95_2} S. Ishida, S. Fujii, S. Kashiwagi, S. Asano, J. Phys.
Soc. Jpn. \textbf{64} (1995) 2152.

\bibitem{Ishida90} S. Fujii, S. Sugimura, S. Ishida, S. Asano, J. Phys.:
Condens. Matter \textbf{2} (1990) 8583.

\bibitem{Noda} Y. Noda, Y. Ishikawa, J. Phys. Soc. Jpn. \textbf{40} (1976)
690.

\bibitem{69W1} P. J. Webster, Contemp. Phys. \textbf{10} (1969) 559.

\bibitem{mnal} Y. Kurtulus, R. Dronskowski, J. Solid State Chem. \textbf{176}
(2003) 390.

\bibitem{ole} O. K. Andersen, Phys. Rev. B \textbf{12} (1975) 3060.

\bibitem{TBLMTO} O. K. Andersen, O. Jepsen, Phys. Rev. Lett. \textbf{53}
(1984) 2571.

\bibitem{vosko} S. H. Vosko, L. Wilk, M. Nusair, Can. J. Phys. \textbf{58},
(1980) 1200.

\bibitem{me1} V. P. Antropov, J. Magn. Magn. Mat. \textbf{262} (2003) L193.

\bibitem{lixt} A. L. Lichtenstein, M. I. Katsnelson, V. A. Gubanov, J.Phys.F 
\textbf{14} (1984) L125.

\bibitem{Antropov} V. P. Antropov, B. N. Harmon, A. N. Smirnov, J. Magn.
Magn. Mat. \textbf{200}, (1999) 148.

\bibitem{chemistry-physics-jargon} Here we stick to the physics jargon in
that ``hybridization'' means what chemists would call ``orbital
interaction'', not to be confused with a unitary transformation into another
set of non-orthogonal but localized orbitals.

\bibitem{landrum-drons} G. A. Landrum, R. Dronskowski, Angew. Chem. Int. Ed. 
\textbf{39} (2000) 1560.

\bibitem{drons-assp} R. Dronskowski, Adv. Solid State Phys. \textbf{42}
(2002) 433.

\bibitem{Schilfgaarde} M. van Schilfgaarde, V. P. Antropov, J. Appl. Phys. 
\textbf{85} (1999) 4827.

\bibitem{AKL} V. P. Antropov, M. I. Katshelson, A. I. Liechtenstein, Physica
B \textbf{237--238} (1997) 336.

\bibitem{Masumoto72} H. Masumoto, K. Watanabe, J. Phys. Soc. Jpn. \textbf{32}
(1972) 281.

\bibitem{Webster} P. J. Webster, K. R. A. Ziebeck, in \textit{Alloys and
Compounds of d-Elements with Main Group Elements}, Part 2, Edited by H. R.
J. Wijn, Landolt-B\"ornstein, New Series, Group III, Vol. 19/c (Springer,
Berlin, 1988), pp. 75--184.
\end{thebibliography}
\end{document}